\documentclass[sigconf]{acmart}

\usepackage{balance}
\usepackage{amsmath}
\usepackage{url}

\usepackage{algorithm}
\usepackage{algpseudocode}
\usepackage{textcomp}
\usepackage{xcolor}

\usepackage{graphicx}
\usepackage{mathtools}
\usepackage{tikz}
\usepackage[export]{adjustbox}
\usepackage{caption}

\usepackage{subcaption}

\usepackage{subcaption}
\usepackage{titlesec}
\usepackage[normalem]{ulem}
\useunder{\uline}{\ul}{}
\usepackage[inline]{enumitem}

\AtBeginDocument{%
  }

\copyrightyear{2025}
\acmYear{2025}
\setcopyright{cc}
\setcctype{by}
\acmConference[SIGIR '25]{Proceedings of the 48th International ACM SIGIR Conference on Research and Development in Information Retrieval}{July 13--18, 2025}{Padua, Italy}
\acmBooktitle{Proceedings of the 48th International ACM SIGIR Conference on Research and Development in Information Retrieval (SIGIR '25), July 13--18, 2025, Padua, Italy}\acmDOI{10.1145/3726302.3730238}
\acmISBN{979-8-4007-1592-1/2025/07}

\begin{document}

\title{PUB: An LLM-Enhanced Personality-Driven User Behaviour Simulator for Recommender System Evaluation}
\renewcommand{\shorttitle}{PUB: Personality-Driven User Behaviour Simulator}


\author{Chenglong Ma}
\email{chenglong.ma@rmit.edu.au}
\orcid{0000-0002-6745-4029}
\affiliation{%
  \institution{Royal Melbourne Institute of Technology}
  \streetaddress{124 La Trobe St}
  \city{Melbourne}
  \country{Australia}
}

\author{Ziqi Xu}
\email{ziqi.xu@rmit.edu.au}
\orcid{0000-0003-1748-5801}
\affiliation{%
  \institution{Royal Melbourne Institute of Technology}
  \streetaddress{124 La Trobe St}
  \city{Melbourne}
  \country{Australia}
}

\author{Yongli Ren}
\email{yongli.ren@rmit.edu.au}
\orcid{0000-0002-3137-9653}
\affiliation{%
  \institution{Royal Melbourne Institute of Technology}
  \streetaddress{124 La Trobe St}
  \city{Melbourne}
  \country{Australia}
}

\author{Danula Hettiachchi}
\email{danula.hettiachchi@rmit.edu.au}
\orcid{0000-0003-3875-5727}
\affiliation{%
  \institution{Royal Melbourne Institute of Technology}
  \streetaddress{124 La Trobe St}
  \city{Melbourne}
  \country{Australia}
}

\author{Jeffrey Chan}
\email{jeffrey.chan@rmit.edu.au}
\orcid{0000-0002-7865-072X}
\affiliation{%
  \institution{Royal Melbourne Institute of Technology}
  \streetaddress{124 La Trobe St}
  \city{Melbourne}
  \country{Australia}
}

\begin{abstract}
    Traditional offline evaluation methods for recommender systems struggle to capture the complexity of modern platforms due to sparse behavioural signals, noisy data, and limited modelling of user personality traits. While simulation frameworks can generate synthetic data to address these gaps, existing methods fail to replicate behavioural diversity, limiting their effectiveness. To overcome these challenges, we propose the \textbf{P}ersonality-driven \textbf{U}ser \textbf{B}ehaviour Simulator (PUB), an LLM-based simulation framework that integrates the Big Five personality traits to model personalised user behaviour. PUB dynamically infers user personality from behavioural logs (e.g., ratings, reviews) and item metadata, then generates synthetic interactions that preserve statistical fidelity to real-world data. Experiments on the Amazon review datasets show that logs generated by PUB closely align with real user behaviour and reveal meaningful associations between personality traits and recommendation outcomes. These results highlight the potential of the personality-driven simulator to advance recommender system evaluation, offering scalable, controllable, high-fidelity alternatives to resource-intensive real-world experiments.\footnote{~The source code can be found at~\url{https://github.com/ChenglongMa/PUB}.}
\end{abstract}



\maketitle
\section{Introduction}
Modern recommender systems (RSs) face critical challenges in evaluation methodologies, particularly as traditional offline datasets struggle to capture the dynamic complexity of user interactions on contemporary platforms. These datasets often lack granular behavioural signals (e.g., personality-driven decision-making) and exhibit biases from sparse or noisy logs, limiting their utility for robust system optimisation. While user studies and real-world datasets provide valuable insights, they are resource-intensive, suffer from uncontrollable confounding variables \cite{ma2022evaluation, zhang2025deconfounding}, and fail to reproduce differentiable evaluation results for modern RSs (e.g., short video platforms) \cite{xu2025irt}. Recent advances in large language models (LLMs) offer promising avenues for simulating user behaviour \cite{wang2024recagent, gao2024survey}, yet existing frameworks remain inadequate in modelling individual differences, such as personality traits, which significantly influence preferences and engagement patterns.  

The integration of personality-driven modelling into RSs \cite{dhelim2023personality} has demonstrated benefits, including improved accuracy and cold-start mitigation. The Big Five personality model \cite{goldberg1992big5,roccas2002big5} (i.e., Openness (O), Conscientiousness (C), Extraversion (E), Agreeableness (A), Neuroticism (N)) has emerged as a gold standard due to its psychological validity and cross-cultural reproducibility. However, simulating these personality traits at scale remains challenging. Prior work in personality computing primarily focuses on inferring traits from static user data (e.g., social media posts \cite{azucar2018footprints,lambiotte2014footprints}), while LLM-based simulations often prioritise generic behavioural patterns over trait-specific dynamics. For instance, research \cite{sorokovikova2024big5,safdari2023llmbig5,jiang2024llmbig5} reveal that LLMs like GPT-4 exhibit distinct Big Five profiles under varying prompts, but their stability and applicability to recommendation tasks remain understudied.  

Existing simulation tools face two key limitations:  
\begin{enumerate*}[label=(\arabic*)]  
  \item \emph{Low fidelity}: Generated behaviours often deviate from real-world statistical distributions, undermining the reliability of recommender system evaluation \cite{ma2022nest, gao2024survey}.  
  \item \emph{Oversimplified personalisation}: Collaborative filtering hybrids or Markov decision processes based methods fail to capture the nuanced correlations between user traits and behaviours \cite{ie2019recsim, wang2024recagent}.  
\end{enumerate*}  
Despite their potential to mitigate ethical and logistical challenges associated with real-user experiments, such as privacy concerns and experimental costs, these limitations hinder the widespread adoption of simulation tools for large-scale recommender system evaluation \cite{wang2024recagent, gao2024survey}.

To address these challenges, we propose \textbf{P}ersonality-driven \textbf{U}ser \textbf{B}ehaviour Simulator (PUB), which synergises LLM-driven personality inference with dynamic behaviour simulation. Our approach is based on two key insights. First, user interactions encode rich psychological signals that serve as proxies for personality traits \cite{tsao2010shopping,lambiotte2014footprints,azucar2018footprints}. PUB leverages these digital footprints along with item metadata to enhance LLM-based trait inference, improving the fidelity of simulated behaviours. Second, recent studies \cite{roccas2002big5,sorokovikova2024big5,jiang2024llmbig5,safdari2023llmbig5} show that LLMs can simulate stable personality profiles when guided by structured prompts. PUB extends this by grounding simulations in empirically validated Big Five correlations, ensuring psychologically coherent behavioural outputs. In summary, our paper makes the following contributions:

\begin{itemize}[leftmargin=0.5cm]
  \item \textbf{Personality-driven simulation}: We introduce a personality-based user behaviour simulation framework that dynamically maps inferred Big Five traits to probabilistic behaviour models.
  \item \textbf{Evaluation robustness}: Experiments show PUB-generated logs can effectively replicate performance trends of various recommendation algorithms when compared to real data.
  \item \textbf{Fine-grained analysis}: Our analysis reveals significant correlations between personality traits and recommendation susceptibility, offering new insights into user behaviour dynamics.
\end{itemize}

\section{Methodology}

\begin{figure}[t]
  \centering
  \includegraphics[width=0.95\columnwidth]{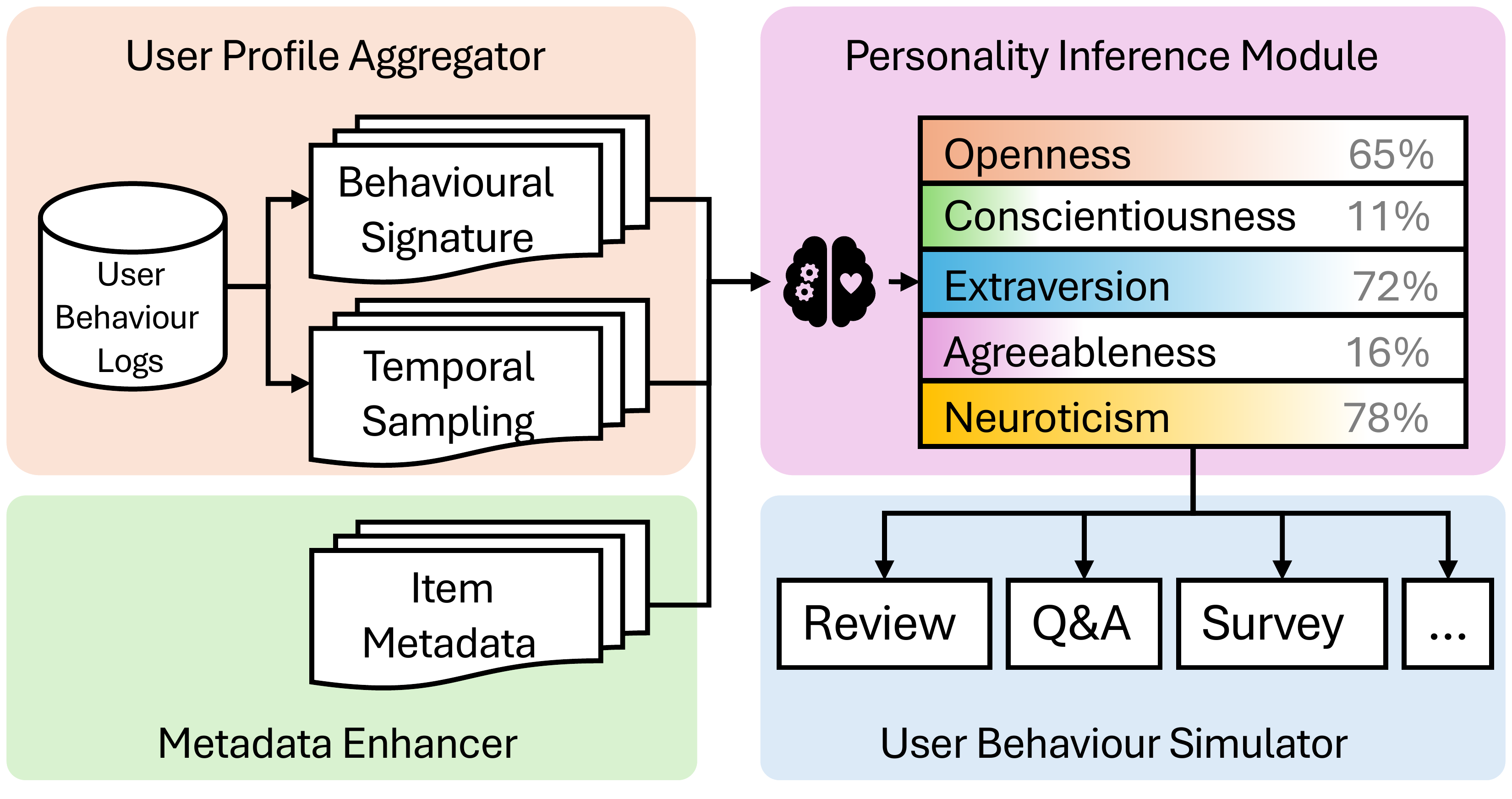}
  \caption{Overview of the proposed PUB architecture.}
  \Description{Overview of the proposed PUB architecture.}
  \label{fig:tube_architecture}
\end{figure}

The proposed PUB is a hybrid architecture designed to simulate personality-driven user behaviour for recommender system evaluation. As illustrated in Fig. \ref{fig:tube_architecture}, the framework operates in four phases:  
\begin{enumerate*}[label=(\arabic*)]  
  \item \textbf{User Profile Aggregator}: Extracts user behavioural logs and computes aggregated features (e.g., purchase frequency, category preferences) using statistical functions.  
  \item \textbf{Metadata Enhancer}: Embeds item metadata (e.g., title, description, price) to enrich contextual information for downstream processing.  
  \item \textbf{Personality Inference Module}: Maps user behaviour logs and item metadata to Big Five personality traits, leveraging prompt-guided LLMs and psychometric mapping.  
  \item \textbf{User Behaviour Simulator}: Generates synthetic interactions conditioned on inferred traits to preserve statistical fidelity to real-world patterns in downstream tasks.  
\end{enumerate*}  

PUB is designed to be dataset-agnostic and applicable across various domains. In the following sections, we use Amazon review datasets \cite{hou2024amazon} as an example.

\subsection{User Profile Aggregator}
The User Profile Aggregator employs a hierarchical approach to construct statistical user profiles from heterogeneous behavioural data. We formalise the process through two primary stages: Behavioural Signature Extraction and Temporal Stratified Sampling.

\textbf{Behavioural Signature Extraction}. The core behavioural patterns are formalised through statistical feature engineering. We first aggregate all category-specific interactions into a unified dataset, denoted as \(\mathcal{D} = \bigcup \left\{\mathcal{D}_c\right\}_{c\in C}\), where \(C\) represents the complete set of Amazon categories. Subsequently, we extract user behavioural logs, comprising sequential interactions \(d = (i_k, t_k, r_k, c_k)\), where \(i_k\) denotes the item ID, \(t_k\) the timestamp, \(r_k\) the rating (1–5 stars), and \(c_k\) the item category. The module computes aggregated features (e.g., purchase frequency, purchase rhythm, category preferences, and price tier distribution) via domain-specific functions.

For example, to model purchase rhythm, we employ circular statistics to capture periodic behavioural patterns based on interaction timestamps. Each interaction timestamp \(t_i\) is mapped onto a unit circle using an angular transformation: $\theta_i = \frac{2\pi t_i}{y \times 24\times60\times60}$, where \(y\) denotes the number of days in the observed cyclic behaviour. This transformation normalises timestamps within a \(2\pi\) range, ensuring that interactions occurring at the same relative position but on different cycles are mapped to similar angular positions. The purchase rhythm is then computed as the mean resultant vector: $\gamma = \frac{1}{n}\sum_{i=1}^n e^{j\theta_i}$, where \(n\) represents the number of interactions. 

The interval entropy measures regularity using Shannon entropy \cite{lin1991entropy} over purchase intervals: $H_{\text{int}} = -\sum_{t\in\mathcal{T}} p(t)\log p(t)$,
where $p(t)$ denotes the probability of purchase interval $t$, and $\mathcal{T}$ represents the set of all intervals~\footnote{~Please refer to our code for other feature definitions.}.

Finally, the module computes user-specific statistical features, denoted as $\mathbf{S}_u = \left\{\gamma, H_{\text{int}}, \dots\right\} \in \mathbb{R}^{d_u}$, where $d_u$ is the feature dimensionality. This module acts as a bridge between raw behavioural logs and trait inference, providing interpretable user representations.

\textbf{Temporal Stratified Sampling}.
Given the longitudinal nature of user interactions, we partition the interaction records into $K$ temporal bins using exponentially increasing time windows. This adaptive sampling strategy maintains chronological integrity while efficiently reducing computational complexity:
\begin{equation}
\small
  \label{eq:temporal_sampling2}
  \mathcal{D}' = \bigcup_{k=1}^{K} \text{Sample}\Big(\mathcal{D}|_{\Delta t_k},\, \eta\Big),
\end{equation}
where $\mathcal{D}|_{\Delta t_k}$ denotes the subset of $\mathcal{D}$ corresponding to the temporal bin of length $\Delta t_k$, $\eta$ represents the sampling length for each bin, and $K$ is the total number of bins.

The adaptive time windows are defined as:
\begin{equation}
\small
\label{eq:temporal_windows}
\Delta t_k =
\begin{cases}
1 \text{ week}, & \text{if } t_{\text{end}} - t_{\text{start}} \leq 1 \text{ year}, \\
1 \text{ month}, & \text{if } 1 \text{ year} < (t_{\text{end}} - t_{\text{start}}) \leq 3 \text{ years}, \\
1 \text{ quarter}, & \text{otherwise}.
\end{cases}
\end{equation}
The thresholds in Eq. \eqref{eq:temporal_windows} are determined empirically.

\subsection{Metadata Enhancer}
The Metadata Enhancer module enriches item metadata (e.g., title, description, price) by incorporating it into LLM prompt-guided contexts. The metadata is then modulated by user-specific statistical features to enhance contextual information. Formally, let $\mathbf{m}_i \in \mathbb{R}^{d_i}$ denote the raw metadata for item $i$, where $i\in \mathcal{D}'$, and let $\mathbf{M}_u = \left[ \mathbf{m}_i\right]_{i \in \mathcal{I}_u}$, where $\mathcal{I}_u$ represents the set of items consumed by user $u$. The enhanced metadata context $\mathbf{C}_{u}$ is then obtained through a prompt-guided fusion function $g(\cdot)$ as $\mathbf{C}_{u} = g\left(\mathbf{M}_u, \phi(\mathbf{S}_u)\right)$,
where $\phi(\cdot)$ transforms statistical features by normalising and scaling them for compatibility with the raw metadata. The function $g(\cdot)$, implemented via an LLM prompt, integrates enriched metadata with transformed statistical cues. The whole process yields a dynamic, task-specific representation $\mathbf{C}_{u}$ that emphasises personality-relevant signals.

\subsection{Personality Inference Module}  
The Personality Inference Module employs a psychometric mapping function to estimate users' Big Five personality traits from the enhanced context: $\mathbf{T}_u = f(\mathcal{C}_u) = \left\{O, C, E, A, N\right\}$. To disentangle the model design from its dependency on specific LLM reasoning abilities \cite{wang2025limits}, we leverage well-established psycholinguistic correlates to guide the trait inference process.  Drawing upon established correlations between digital footprints and psychological constructs \cite{goldberg1992big5,tsao2010shopping,lambiotte2014footprints,azucar2018footprints}, we formalise this mapping as follows:
\begin{itemize}[leftmargin=0.6cm]
    \item \textbf{Openness}: Category entropy \cite{tsao2010shopping} and metaphor density.
    \item \textbf{Conscientiousness}: Review length consistency, rating deviation from category averages, and purchase rhythm regularity.
    \item \textbf{Extraversion}: Social reference frequency (e.g., ``we'', ``gift'' via the LIWC-22 lexicon \cite{boyd2022liwc,goldberg1981lexicons}).
    \item \textbf{Agreeableness}: Positive sentiment ratio (e.g., VADER \cite{hutto2014vader}) and politeness markers \cite{goldberg1981lexicons}.
    \item \textbf{Neuroticism}: Negative emotion volatility \cite{goldberg1981lexicons}.
\end{itemize}

The prompt is structured, to encapsulate the user's behavioural statistics and metadata context. The response contains the inferred trait scores, which subsequently guide the behaviour simulation process, enabling diverse and personalised interactions.

\subsection{User Behaviour Simulator}
The User Behaviour Simulator generates synthetic interactions conditioned on inferred trait distributions, preserving statistical fidelity to real-world patterns. Based on the inferred personality, the LLM-based agent performs various downstream tasks, such as completing Q\&A tasks, and providing feedback on recommendations. In this study, we focus on the recommendation task, where the agent generates synthetic interactions. These interactions are then evaluated against real-world data using standard metrics (e.g., nDCG). While our scope is limited to this application, the proposed PUB framework is broadly applicable and can be extended to other domains.

\section{Experiments}
\subsection{Experimental Setup}
Our experiments use the Amazon Review Datasets \cite{hou2024amazon}, which contains 571.54 million interactions across 30 distinct categories from 1996 to 2023. To ensure methodological rigour, we apply a three-stage preprocessing pipeline informed by established practices in sparse recommender systems \cite{singh2020scalability}. First, interactions across all categories are aggregated to model cross-domain behavioural diversity. Second, users and items with fewer than 20 interactions are filtered to reduce sparsity-induced noise. Finally, to preserve temporal dynamics \cite{ma2024temporal}, the dataset is chronologically partitioned into training set $\mathcal{D}_{\text{tr}}$ and testing set $\mathcal{D}_{\text{te}}$.

\begin{figure}[t]
    \centering
    \begin{subfigure}[b]{0.49\linewidth}
        \centering
        \includegraphics[width=\linewidth]{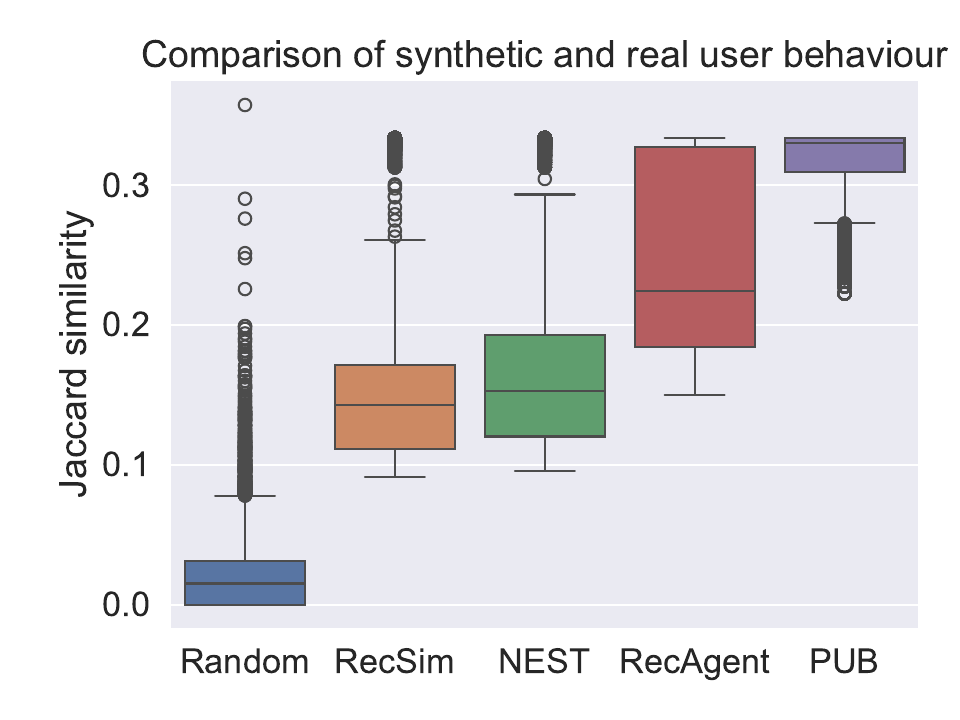}
        \caption{}
        \label{fig:synthetic_real_comparison}
    \end{subfigure}
    \begin{subfigure}[b]{0.49\linewidth}
        \centering
        \includegraphics[width=\linewidth]{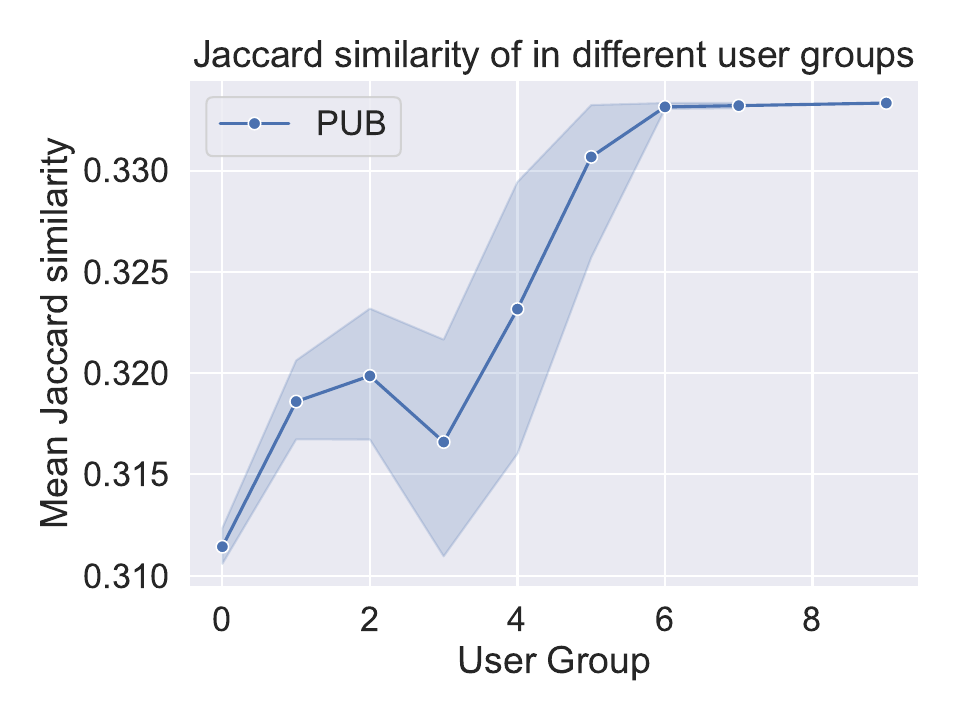}
        \caption{}
        \label{fig:pub_user_group}
    \end{subfigure}
    \caption{(a) Comparison of synthetic and real user behaviour sequences; (b) Jaccard similarity of different user groups.}
    \label{Fig:metaphor_examples}
    \Description{(a) Comparison of synthetic and real user behaviour sequences; (b) Jaccard similarity of different user groups.}
\end{figure}

\begin{figure*}[hptb]
    \centering
    \begin{subfigure}[b]{0.33\linewidth}
        \centering
        \includegraphics[width=\linewidth]{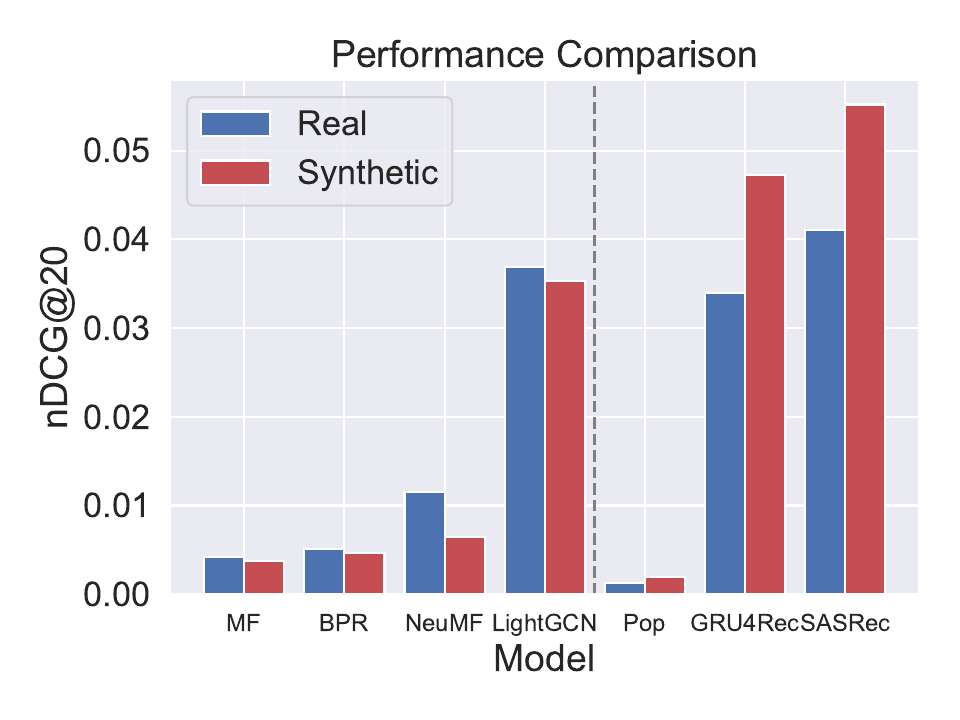}
        \caption{}
        \label{fig:algorithm_performance}
    \end{subfigure}
    \begin{subfigure}[b]{0.33\linewidth}
        \centering
        \includegraphics[width=\linewidth]{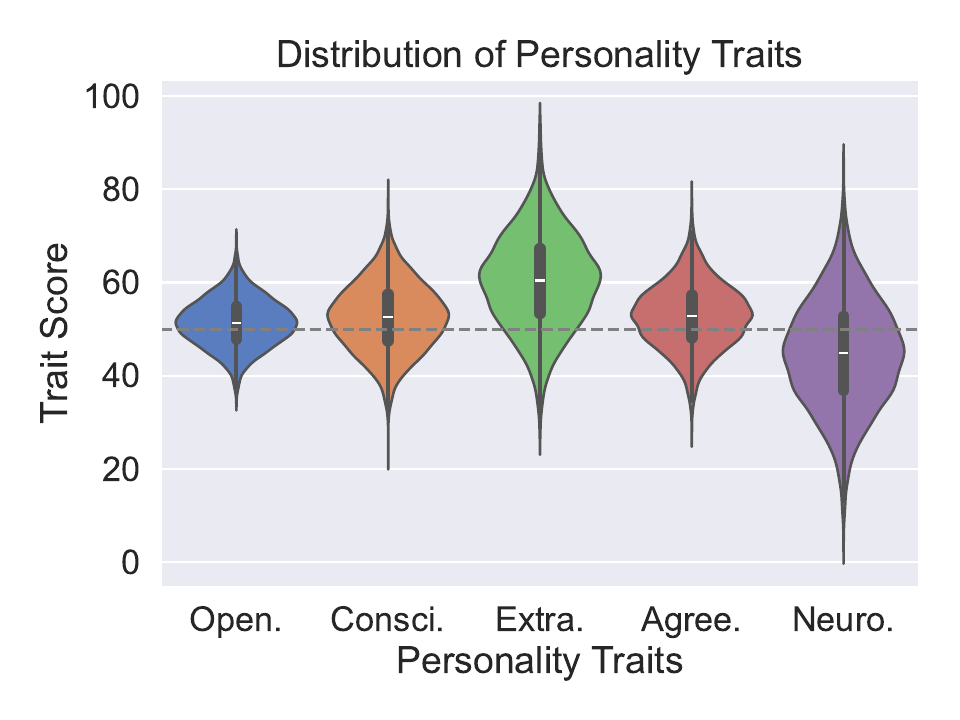}
        \caption{}
        \label{fig:personality_distribution}
    \end{subfigure}
    \begin{subfigure}[b]{0.33\linewidth}
        \centering
        \includegraphics[width=\linewidth]{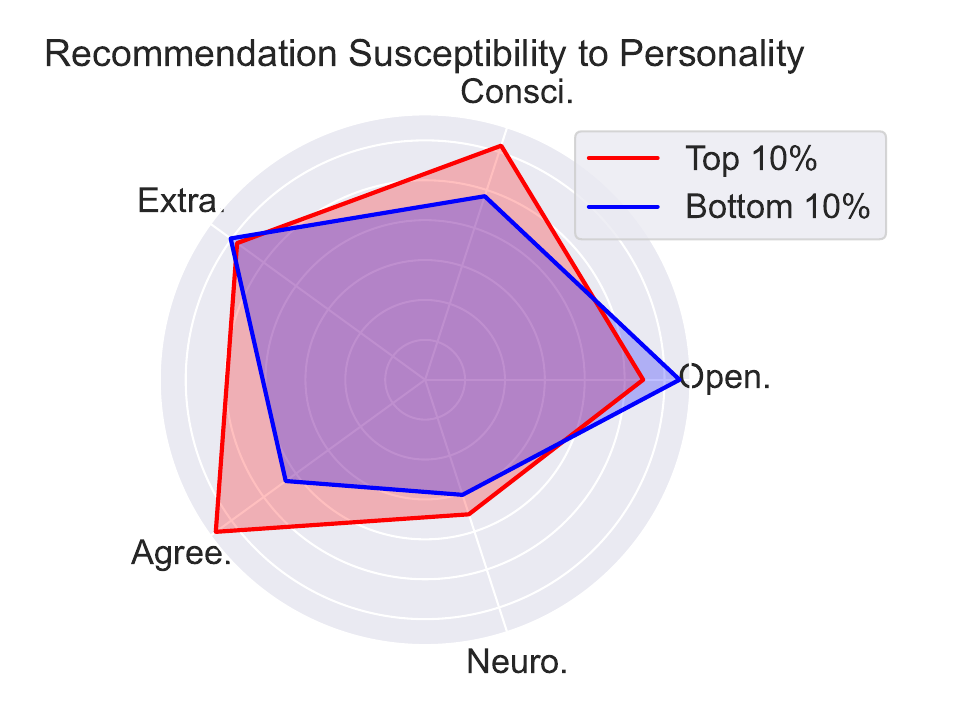}
        \caption{}
        \label{fig:personality_susceptibility}
    \end{subfigure}
    \caption{(a) Performance comparison; (b) Distribution of personality traits; (c) Recommendation susceptibility to personality.}
    \label{}
    \Description{(a) Performance comparison; (b) Distribution of personality traits; (c) Recommendation susceptibility to personality.}
\end{figure*}

\subsection{Research Questions and Analysis}

\subsubsection{\textbf{RQ 1:} Can the proposed PUB generate synthetic behaviour sequences that closely resemble real user interactions?}

Building on \cite{wang2024recagent}, we derive each user's initial personality distribution and shopping behaviour patterns from their interaction history $\mathcal{I}_u$ in the training dataset $\mathcal{D}_{\text{tr}}$. The test set interactions $\mathcal{P}_u$ serve as the ground truth. To isolate the effects of the recommendation algorithm, we employ a proxy recommender. At each iteration $t$, this model constructs a mock recommendation list $\mathcal{L}_t$ for user $u$:
$\mathcal{L}_t = \{i_p, i_{n1}, i_{n2}, \dots, i_{nk}\}, k=9$. Here, $i_p$ is the positive sample, chronologically drawn from $\mathcal{P}_u$, while each $i_{nk}$ is a negative sample randomly selected from $\mathcal{I}_u^- = \{i \mid i \notin \mathcal{I}_u\}$.  The user agent then selects the most relevant item from $\mathcal{L}_t$ based on its inferred personality, forming the synthetic sequence $\mathcal{S}_u$. We measure the Jaccard similarity \cite{bag2019jaccard} between $\mathcal{S}_u$ and the true interaction sequence $\mathcal{P}_u$, where a higher value indicates better alignment with real behaviour.

We compare PUB with four baseline simulation frameworks: \emph{Random Sampling}, which randomly selects items without user preferences; \emph{RecSim} \cite{ie2019recsim}, a Markov Chain-based simulation; \emph{NEST} \cite{ma2022nest}, an extension of RecSim incorporating user need states; \emph{RecAgent} \cite{wang2024recagent}, an LLM-based simulation that randomly assigns user profiles (e.g., age, gender, occupation).  

The results in Fig. \ref{fig:synthetic_real_comparison} show that PUB-generated sequences align closely with real user interactions, achieving an average Jaccard similarity of 0.31. While RecAgent performs well in top cases, its overall performance is less stable. These findings suggest that PUB effectively replicates real-world user behaviour, providing a reliable foundation for further evaluation. 
To further assess the impact of interaction frequency, we group users into 10 clusters (0 is the least frequent and 9 is the most frequent) and measure similarity for each. Fig. \ref{fig:pub_user_group} shows that Jaccard similarity increases with interaction frequency, indicating that richer interaction histories contribute to more accurate personality modelling.

\subsubsection{\textbf{RQ 2:} Can the proposed PUB accurately evaluate the performance of different recommendation algorithms?}

In this experiment, synthetic user behaviour sequences are generated following the aforementioned methodology and then divided into a training set, $\mathcal{S}_{\text{tr}}$, and a test set, $\mathcal{S}_{\text{te}}$. To ensure a fair comparison, the synthetic test set is matched in size to the original test data. A variety of recommendation algorithms are evaluated, including \emph{Pop}, \emph{MF} \cite{koren2009mf}, \emph{BPR} \cite{rendle2012bpr}, \emph{NeuMF} \cite{he2017neural}, \emph{LightGCN} \cite{he2020lightgcn}, \emph{GRU4Rec} \cite{hidasi2015gru4rec}, and \emph{SASRec} \cite{kang2018sasrec}. Their performance is assessed using ranking metrics (e.g., nDCG@20) on both real and synthetic test sets.

As shown in Fig. \ref{fig:algorithm_performance}, the performance of each algorithm on the synthetic test set closely mirrors that on the real test set, suggesting that the proposed simulation model is a viable alternative for evaluating recommendation algorithms. Notably, MF, BPR, NeuMF, and LightGCN generally perform worse on the synthetic test set, while Pop, GRU4Rec, and SASRec perform better. The former group primarily relies on collaborative filtering, leveraging the user-item interaction matrix, whereas the latter (except Pop) focuses on sequential recommendation, capturing sequential or attentional patterns in user interactions.

This discrepancy likely arises because the synthetic data is generated based on users' inferred \emph{personality traits} rather than the \emph{collaborative behaviour} observed in real-world data. Pop achieves higher performance on the synthetic test set as PUB links users' personality traits with their preference for item popularity during profile construction, aligning with the recommendation strategy of Pop. These findings suggest a potential enhancement for our simulation framework by incorporating social influences and collaborative signals among users. 

\subsubsection{\textbf{RQ 3:} What are the distributions of the Big Five personality traits, and how do these traits relate to shopping behaviour?}

An association analysis seeks to determine which personality traits are more indicative of users who are prone to leaving detailed shopping traces and reviews. 

As shown in Fig. \ref{fig:personality_distribution}, the distribution of the Big Five personality traits within the Amazon platform appears relatively balanced, with Extraversion being the most prevalent trait. Notably, the average Neuroticism score is significantly lower than that of other traits, suggesting that users who leave reviews on Amazon tend to be less emotionally stable. This aligns with the hypothesis that writing a review represents a stronger interaction signal than making a purchase, as it requires greater effort and emotional investment. Users with \emph{extreme} purchase experiences are more likely to leave reviews and express emotions in their feedback, potentially correlated with lower neuroticism scores. In contrast, users with \emph{moderate} purchase experiences may be more inclined to provide ratings or refrain from leaving feedback altogether, making it more difficult to infer their personality traits. This poses a challenge for recommender systems in ensuring fair, consistent, and high-quality recommendations.

\subsubsection{\textbf{RQ 4:} Which personality traits are favoured by recommendation algorithm?}

We investigate the relationship between personality traits and recommendation outcomes by selecting users with extreme performance (top 10\% and bottom 10\% nDCG@20). Note that we employ GRU4Rec as the recommender algorithm. As shown in Fig. \ref{fig:personality_susceptibility}, users with high Agreeableness and Conscientiousness scores tend to receive better recommendations, whereas, surprisingly, those with high Openness scores exhibit lower accuracy. This suggests that users who are more open to new experiences may disrupt the algorithm's inference process by deviating from established behavioural patterns. In contrast, individuals with higher levels of agreeableness and conscientiousness are more likely to adhere to recommendations and provide detailed feedback, thereby enhancing the system's performance.

\section{Conclusion}

In this paper, we propose the Personality‐driven User Behaviour Simulator, which embeds the Big Five personality traits into user‐behaviour modelling. Our evaluation demonstrates that PUB's synthetic interaction logs closely mirror real user data, yielding a scalable and reliable platform for recommender system evaluation. Future work will focus on integrating social influence to further enrich the framework.


\section{Acknowledgements}
This research was conducted by the ARC Centre of Excellence for Automated Decision-Making and Society (ADM+S, CE200100005), and funded fully by the Australian Government through the Australian Research Council.

\bibliographystyle{ACM-Reference-Format}
\balance
\bibliography{references}

\end{document}